\documentclass[preprint,aps,showkeys,showpacs,preprintnumbers,amsmath,amssymb]{revtex4}

\usepackage{graphicx}
\usepackage{dcolumn}
\usepackage{bm}

\begin{document}


\title{Role of heterostructures and multiple magnetic phases in 
the low-field magnetization of Fe-Cr GMR multilayers}

\author{R. S. Patel}
\author{A. K. Majumdar}
 \email{akm@iitk.ac.in}
\affiliation{Department of Physics, Indian Institute of Technology,
Kanpur-208 016, India}

\author{A. K. Nigam}
\affiliation{Tata Institute of Fundamental Research, Homi Bhabha Road,
Mumbai-400 005, India}

\author{D. Temple}
\author{C. Pace}
\affiliation{MCNC, Electronic Technologies Division,
Research Triangle Park, North Carolina, 27709}


\begin{abstract}

Zero-field-cooled (ZFC) and field-cooled (FC) magnetization
along with ac-magnetization vs. temperature
and m-h loop measurements are reported 
for two series of ion-beam sputtered Fe-Cr GMR multilayers
where the interface roughness is different. The exchange 
coupling between the Fe layers varies from ferromagnetic (FC) 
to antiferromagnetic (AF) depending upon the Cr layer
thickness. The ZFC and FC magnetization data 
follow different curves below an irreversible temperature 
($T_{irr}$). The FC data shows a $T^{3/2}$ thermal demagnetization
behavior at lower temperatures with very small spin-wave stiffness constant 
(as compared with that of bulk Fe obtained from 
Bloch's $T^{3/2}$ law) but it goes as 1/T at higher temperatures (above ($T_{irr}$)). 
This behavior is interpreted in terms of the coexistence of spin-glass
(SG)/superparamagnetic, FM and AF phases. ac-magnetization 
vs. temperature shows a peak at $T_g$. This peak
shifts towards higher temperatures and its amplitude decreases
with increasing frequency of the driving ac field.
\end{abstract}

                             

\maketitle

\section{INTRODUCTION}
The giant magnetoresistive (GMR) property was discovered in 1988 \cite{Baibich:1988}. Since then
it has become an established area of promising technological applications. Sensors
and read heads in hard disks are major attractions of GMR materials. Tera byte
capacity hard disks are not far from coming to the market. In the present 
Fe-Cr GMR materials ferromagnetic (FM)
Fe layers are stacked antiferromagnetically with non-magnetic Cr spacer layers.
This antiferromagnetic arrangement of the Fe layers is engineered by varying the Cr
spacer layer thickness. With varying Cr thickness successive Fe layers show
oscillatory antiferromagnetic (AF) and FM couplings.
In an external magnetic field $H>H_{sat}$, the Fe layers
of these Fe-Cr samples align ferromagnetically whereas in zero field they are in 
antiferromagnetic configuration.
Our study is focused on finding out the magnetic behavior of multilayers in low
external magnetic fields (up to a few hundred gauss). The low field response
of multilayers is important for technical applications as well as for interesting
physics behind it.

Kravtsov et al. \cite{Kravtsov:2001} studied the interface formation and magnetic
ordering in Fe-Cr multilayers with Fe thickness 2, 4, and 13 \AA \ by polarised
neutron reflectometry. They found that the samples with Fe layer thickness
of 2 \AA \ show pure superparamgnetic behavior. Increasing the Fe thickness
to 4 \AA \ led to a crossover from superparamagnetic to a mixed ferromagnetic
and spin-glass state. The sample with 13 \AA \ Fe layer displays the usual magnetic
properties of a GMR multilayer. Here one should keep in mind that 2 \AA \ is
too thin for a layer which will rather be a discontinuous magnetic layer. Fert et al. \cite{Fert:1995}
referred such type of multilayers as hybrid nanostructured multilayers because 
they consist of both usual layers and layers having nano-scale lateral structure,
like clusters. Low-field magnetization study of CoFe-Al$_2$O$_3$ multilayers by
Kakazei et al. \cite{Kakazei:2003} showed that depending upon the CoFe thickness
and temperature, the system shows superparamgnetic, spin-glass or a FM like state.
They also concluded a co-existence of different phases in the mixed state. An 
experimental investigation of the magnetic properties of multilayer (Gd/Si/Co/Si)$_n$
films in low magnetic fields by Patrin et al. \cite{Patrin:2003} had shown a spin-glass
like behavior. They explained the existence of the spin-glass state of multilayer
films in terms of the bi-quadratic exchange interaction. de Oliveira et
al. \cite{Oliveira:1999} explained the magnetic irreversibility in Fe-Cu multilayers
at low temperatures in terms of the presence of some interdiffusion between 
Fe and Cu at the interfaces.
They also found diffused Fe atoms aggregate in clusters at the interfaces.
The magnetization behavior of thin (10 - 50 \AA) epitaxial Fe films on Cr
studied by Berger and Hopster \cite{Berger:1994} by the magneto-optical 
Kerr effect revealed that the exchange coupling between Cr and Fe overlayers
depends very much on the intrinsic antiferromagnetic properties of Cr. At 123 K, Cr
shows a phase transition from longitudinal spin density wave (SDW) to
transverse SDW.

\section{EXPERIMENTAL DETAILS}
Our samples are grown on Si substrate by ion beam sputter deposition
technique using Xe ion at 900 V with a beam current of 20 mA and 1100 V 
with a beam current of 30 mA. The typical
structures are Si/Cr(50 \AA)/[Fe(20 \AA)/Cr(t \AA)]$\times$ 30/Cr(50 -t \AA).
Sample 1 has t = 10  \AA \ and samples 2 and 3 have t = 12  \AA \ but they are
deposited under different base pressure at 900 V. Different base pressure results
in different surface roughness. Let us call these samples 1 - 3,
{\it series A} samples. Samples 4 - 8 have t = 6, 8, 10, 12, and 14 \AA, \ respectively 
and they are sputtered at 1100 V. Let us call these samples 4 - 8, {\it series B} samples.
These samples 
are well characterised and the details have been given elsewhere
\cite{Lanon:2002, Majumdar:2002}.
All the experiments were done with a Quantum Design superconducting 
quantum interference device (SQUID) magnetometer
(MPMS). The magnetic field is applied in the plane of the multilayer samples.
Samples 1 - 8 have GMR $(=(\rho(H)-\rho(0))/\rho(0) \times 100 \%)$ of $\sim$ 20, 21, 21,
0.4, 31, 33, 32, and 29 \%, respectively at 4.2 K in a longitudinal magnetic field $\sim$ 1 tesla.
ac-magnetization was done using a Quantum Design PPMS with some
modifications at Forschungszentrum Karlsruhe, Germany.

\section{RESULTS AND DISCUSSION}
\subsection{dc-magnetization}
\begin{figure}
\includegraphics[bb=0.4cm 0.5cm 8.7cm 7cm,width=8.5cm]{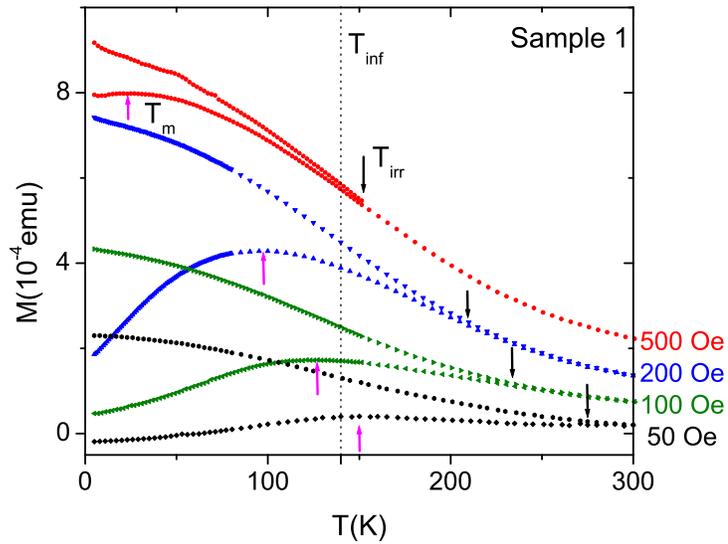}
\caption{\label{fig:s1raw} (Color online) Magnetization {\it vs.} temperature (raw data without any
corrections for the diamagnetism of Si substrate and paramagnetism of the packing materials) for sample 1.
Three characterstic temperatures $T_m$(marked with $\uparrow$), $T_{irr}$(marked with
$\downarrow$), and $T_{inf}$ (dashed line) are also presented. $T_m$ and $T_{irr}$ decrease
to lower temperatures with increasing applied external magnetic field. $T_{irr}$ is found to be 
independent of the applied external magnetic field. Samples 2, 3, 5, and 6 also show this type of
history-dependent magnetization behaviour.}
\end{figure}

When we cool the Fe-Cr multilayer samples from room temperature to a lower temperature
of interest in zero magnetic field and then apply a small magnetic field in the plane
of the multilayer, moments in iron layers start responding to this external magnetic 
field. Those moments of Fe layers, which are not aligned in the direction of external 
field try to align in the direction of the applied magnetic field. This gives a finite moment
for the whole sample. Good GMR multilayer systems are AF in the sense
that the FM Fe layers are coupled antiferromagnetically in zero field. When we increase the 
temperature from the lowest temperature, the thermal energy starts disrupting this
nearly perfect AF alignment. Thermal fluctuations equally affect both types of Fe layers, i.e.,
the Fe layers with magnetic moments aligned parallel to the applied magnetic field and 
those with magnetic moments aligned antiparallel to the field. However, the
external magnetic field will try to suppress the fluctuations of moments which
are in the direction of the field and will effectively increase the fluctuations of moments which are
aligned in the opposite direction. In other words, the decrease in the magnetization due to the 
thermal energy is less for Fe moments parallel to the applied field. Thus the magnetization 
increases with the increasing temperature. But when the temperature is beyond a 
critical value, the thermal energy starts disrupting all the ordered moments thus 
decreasing the net magnetization with increasing temperature. So for a small 
external magnetic field we observe a peak temperature $T_m$ as 
shown in Fig. \ref{fig:s1raw} for sample 1. This behavior has been seen prominently
in samples 1, 2, 3, 5, and 6 and presented in Table \ref{tab:dct}. This peak temperature is 
a function of the applied magnetic field.

If we cool the sample in the presence of a small magnetic field, we do not find any peak 
in M(T), rather M increases at lower temperatures (more ordered state).
This feature is similar to the history-dependent effect of spin-glass systems 
where zero-field cooled (ZFC) and field-cooled (FC) magnetization curves follow 
different paths at low temperatures and low fields. The motivation behind the present 
study is to explore such behavior at low fields in these GMR multilayers. $T_m$ (which approaches the 
glass transition temperature $T_g$ when $H \rightarrow 0$) decreases with the increasing 
external magnetic field because a higher magnetic field is able to align the fluctuating 
moments to their maximum value at lower temperatures. In the high-temperature limit ZFC and FC 
magnetizations show the same temperature dependence. A characterstic temperature 
$T_{irr}$ can be defined below which the sample shows such history dependent effects. 
This temperature also decreases with increasing field and $T_{irr}$ is 
always greater than $T_m$. These samples show another characterstic temperature, namely, a point of
inflection $T_{inf}$ in FC magnetization vs. T curve. Below this temperature, 
FC curves are convex upwards
and above it they are concave upwards. Thus one more characterstic temperature 
$T_{inf}$ can be derived apart from $T_m$ and $T_{irr}$. This $T_{inf}$ is 
found to be independent of the external magnetic field as shown in Table \ref{tab:dct}.
Similar to this work Durand et al. \cite{Durand:1993} found different 
magnetization in FC and ZFC measurements in Fe-Cu multilayers at low temperatures. 
They interpreted this phenomena by interdiffusion of Fe atoms in Cu
layers. They also found a very low value of the saturation magnetization,
nearly one order of magnitude smaller than that of bcc Fe.

\begin{figure}
\includegraphics[bb=14pt 16pt 271pt 220pt,width=246pt]{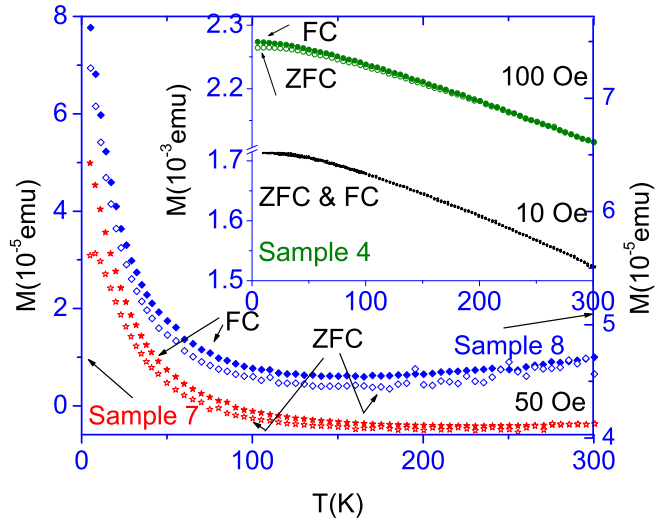}
\caption{\label{fig:ox3raw} (Color online) Magnetization {\it vs.} temperature (raw data without any
corrections for the diamagnetic Si substrate and the paramagnetic packing materials) for sample 
7, 8 and 4 (in the inset).}
\end{figure}

Sample 4 did not show any $T_m$  as shown in the inset of Fig.~\ref{fig:ox3raw}.
It has a small Cr thickness ($t_{Cr}$) of 6 \AA \ which leads to a FM coupling between the Fe layers
and not an AF coupling. This fact is supported by a GMR of less than 1 \% for this sample.
For the same external magnetic field, it can be seen from Table~\ref{tab:dct} that 
$T_m$ shifts to lower temperatures from 75 to 20 K for sample 5 ($t_{Cr}$ = 8 \AA) to sample 
6 ($t_{Cr}$ = 10 \AA) for an applied field of 50 Oe. This suggests that $T_m$
may be at much lower temperatures for samples 7 and 8. Our measurement
temperature range is 5 to 300 K. Sample 7 did not show any peak in a magnetic 
field of 200 Oe, however, in a magnetic field of 50 Oe there is a signature of a peak
at 8 K (notice the first 3-4 points of ZFC curve of sample 7 in Fig.~\ref{fig:ox3raw}). 
Sample 8 showed a totally different behavior at higher temperatures 
compared to the other samples in this series as shown in Fig.~\ref{fig:ox3raw}.
Here the magnetization increases with increasing temperature above 150 K
which could not be understood.

\begin{figure}
\includegraphics[bb=0.4cm 0.6cm 9.5cm 7.6cm,width=8.5cm]{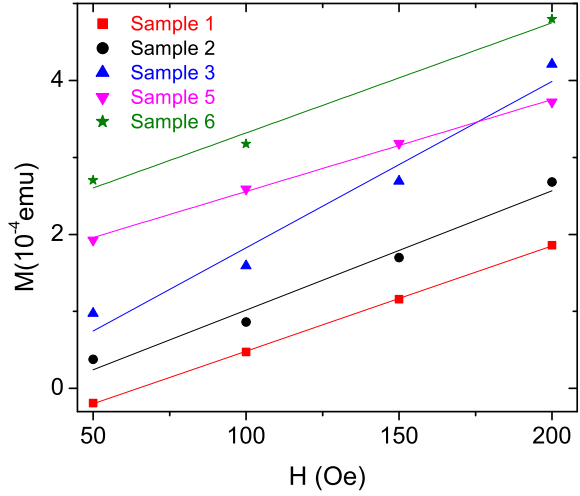}
\caption{\label{fig:gm0ZFCall} (Color online) $m(5 K)_{ZFC}$ {\it vs.} applied external magnetic field for samples 1, 2,
3, 5 and 6. $m(5 K)_{ZFC}$ increases almost linearly with the applied magnetic field. Data and
fit for sample 5 and 6 are shifted along y-axis by adding 1.5 $\times 10^{-4}$ emu and 2 $\times 10^{-4}$
emu, respectively for clarity.}
\end{figure}

If we take the magnetization value ($m(5 K)_{ZFC}$) of the samples at the lowest temperature
from the ZFC magnetization measurements at different applied external magnetic fields, then 
we find that ($m(5 K)_{ZFC}$) is roughly linear with the applied field as in a 
paramagnet. This has been shown in Fig. \ref{fig:gm0ZFCall}.

From experimental point of view some major signatures of spin glasses are: (i) below $T_g$ 
spin-glass systems show history-dependent behavior, i.e., the magnetization
measured in a field-cooled condition is different from that under the zero-field-cooled 
condition \cite{Chowdhury:1986} and (ii) low field, low frequency ac susceptibility $(\chi_{ac}(T))$ 
exhibits a cusp at a temperature $T_g$. The origin of the spin-glass-like behavior here
may be the interface roughness and interdiffused clusters. 
These are sputtered samples and the surface roughness
may lead to Fe-Fe, Fe-Cr and/or Cr-Cr frustrations \cite{Berger:1994, Pierce:1999}.
There may be Fe clusters interdiffused inside Cr layers and vice versa.
In spin glasses, $T_g$ is the temperature below which the spins are frozen, i. e.,
the FC magnetization curve is flat below $T_g$. However, in multilayers, in low fields
AF Cr's role may be quite complex.  Bulk Cr has a N\'{e}el temperature $T_N$ = 311 K. 
N\'{e}el temperature of Cr layers in these multilayers can be very much different from
that of the bulk. It is known that the Curie temperature of Fe films is less than
that of bulk Fe. In Cr, as the thickness decreases, the N\'{e}el temperature also
decreases, either due to the decoupled AF state behaving like a thin film, or
due to the increasing spin frustration due to closer interfaces. The mechanism of 
antiferromagnetism in Cr is well explained by spin density wave 
(SDW) \cite{Berger:1994, Overhausser:1960}. Below
$T_N$, the SDW is transverse, i.e., the magnetic moments are perpendicular to the 
SDW wavevector. This transverse SDW state shows a transition to longitudinal 
SDW state at the spin-flip transition temperature $T_{SF}$ = 123 K \cite{Berger:1994}.
This adds to the difficulty in understanding the role of Cr in the magnetization
study of these multilayers. The magnetization behavior of thin Fe films on Cr
studied by Berger et al. \cite{Berger:1994} by magneto-optical Kerr effect gave indication that
the magnetization in Fe layers is oriented in the Fe-Cr film plane. In the real case, ion
beam sputtered layers may contain atomic steps at surfaces. These atomic steps 
cause frustration while aligning the magnetic moments at the interface. Figure 5 of 
ref. \onlinecite{Berger:1994} and Fig. 3 of ref. \onlinecite{Pierce:1999} give beautiful illustrations
of different atomic interface situations. In summary, we can say that in these 
measurements we get contributions from many sub-systems like FM
Fe films, AF Cr films which show SDW transition at $\sim$ 123 K (below this 
temperature Cr may lead to the formation of small domains in Fe films), and SG-like interdiffused 
clusters, spin waves, etc.

\begin{figure}
\includegraphics[bb=0.4cm 0.5cm 10.5cm 7.8cm,width=8.5cm]{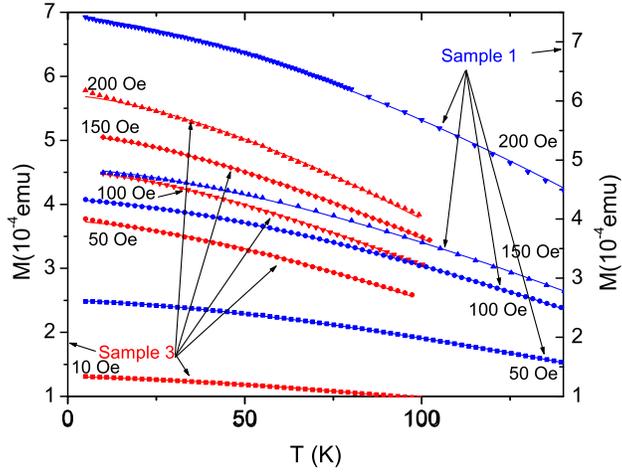}
\caption{\label{fig:3by2} (Color online) Magnetization {\it vs.} temperature for samples 1 and 3.
The solid lines are the fits to Eq. (\ref{eq:pwr}). Data are fitted below $T_{inf}$.}
\end{figure}

The antiferromagnetic exchange coupling between two successive Fe layers
decreases with increasing temperature. Most of the theories assume that the coupling involves
only those few atomic layers in the FM which are nearest to the 
interfaces\cite{Heinrich:1994, Cullen:1993}. Experiments 
introducing other FM materials at the interfaces have confirmed this point of view to some extent.
Each FM layer can be divided into ``bulk" and ``surface" regions consisting of 
superparamagnetic and/or spin-glass particles/clusters, etc.. 
We find that the FC magnetization data of our multilayers fit well to 
the power law of the form

\begin{equation}
\frac{\Delta M}{M(0)} = \frac{M(T)-M(0)}{M(0)}=- A T^{3/2},
\label{eq:pwr}
\end{equation}
\noindent
where A is a constant of proportionality. The values of $\chi^2$ are 
consistent with the experimental
resolution. The values of $\chi^2, R^2$(correlation coefficient),$ M(0),$ and A are presented in Table \ref{tab:pwr}.
The typical fits are presented in Fig. \ref{fig:3by2}. The proportionality constant A,
which is analogous to the the spin-wave stiffness constant in Bloch's $T^{3/2}$ law,
depends on the magnetization. We have earlier done magnetization measurements 
(above H$_{sat}$ where all Fe layers are aligned in FM configuration) with temperature
in these multilayers \cite{Patel:2005}. We fitted the data to Bloch's $T^{3/2}$ law.
We found that the values of A are of the order of $1\times10^{-5}$
K$^{-3/2}$. In low fields, on the other hand, A is 30 times more compared to the above 
value. To explain this 
we have to consider the behavior of Fe clusters in Cr layers. It has been found by Vega et al. 
\cite{Vega:1994} that Fe$_N$ ($N$ = number of Fe atoms in Fe clusters) clusters can behave as a FM or an AF. 
The magnetic order within the Fe and Cr slabs is qualitatively the same as in the
corresponding pure solids and that the coupling at the interfaces is usually AF.
They found strong environment dependence of the local magnetic moments.
The Fe moments at the interface are reduced with respect to the bulk value,
while in the middle of the slab they are sometimes enhanced. The Cr moments
are not only modified close to the interface, but they are extremely sensitive to 
the compatibility of their spin density wave (SDW) state and the AF coupling
at the Fe-Cr interface. They have calculated that small Fe$_N$ clusters in Cr
order AF as the Cr matrix for $N\le4$. For larger Fe$_N (N \ge 6)$, the magnetic 
order within the Fe cluster is FM-like. So `A' value will depend on the behavior
of these Fe clusters. Large values of A will give small spin-wave stiffness constant.
Similar results were found by de Oliveira et al. on Fe-Cu multilayers. Sample 4
gives good fit to Eq.~\ref{eq:pwr} for the whole temperature range and in both
ZFC and FC cases. Here the value of the coefficient A is close to that given in 
ref. \onlinecite{Patel:2005}. This means that the magnetization behavior
of sample 4 even in low fields is more like the magnetization behavior of 
multilayers in the
presence of magnetic filed greater than saturation field.

\begin{figure}
\includegraphics[bb=0.4cm 0.5cm 10.5cm 7.8cm,width=8.5cm]{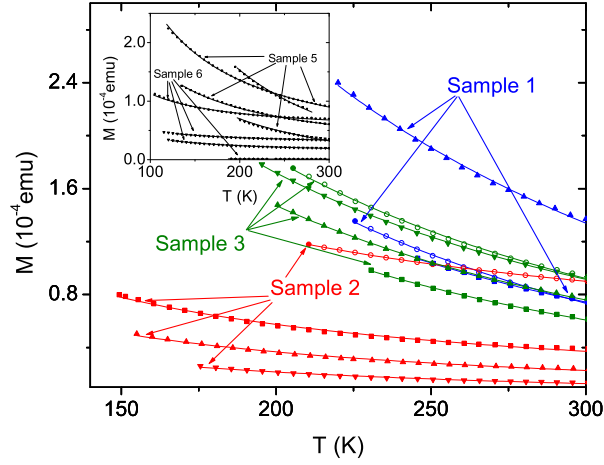}
\caption{\label{fig:onebyT} (Color online) Magnetization {\it vs.} temperature for samples 1, 2, 3
5, and 6 above $T_{irr}$. The solid lines are the fits to an equation of the form $M \sim 1/T$.}
\end{figure}

Above $T_{irr}$ the magnetization decreases as 1/T in these multilayers as shown in Fig.
\ref{fig:onebyT}. At high temperature the thermal energy disrupts all magnetic alignments. 
Similar studies of low-field magnetization in ion-beam deposited
metal-insulator $CoFe-Al_2O_3$ multilayers were carried out by Kakazei et al.
\cite{Kakazei:2003}. They also reported that the low-field magnetization curve
M(H,T) generally displays the Curie-Weiss behavior at high temperatures 
and below a certain blocking temperature the FC-ZFC curves split.
Figure \ref{fig:onebyT} shows the typical fits to $M \sim 1/T$. 
We found that the $R^2 > 0.99$. We 
have taken the data only above $T_{irr}$ for the above analysis.

\begin{figure}
\includegraphics[bb=0.4cm 0.5cm 9.5cm 7.2cm,width=8.5cm]{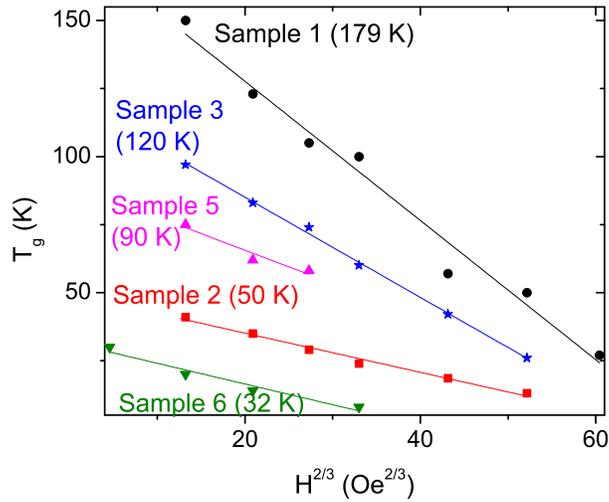}
\caption{\label{fig:at} (Color online) The plot of $T_m$ vs. $H^{2/3}$.
$T_m$-axis intersection gives the $T_g$ for that sample (written in bracket).}
\end{figure}

Now we proceed to find out the external field
dependence of $T_m$. We find that the de Almeida and Thouless (AT) behavior
of the form $H/T \propto (T_g/T-1)^{3/2}$, where $T_g$ is the spin-glass
temperature, gives an unique $T_g$ for each sample which is 
$\sim (1.2-1.5) \times T_{inf}$ of that sample.
Fig.~\ref{fig:at} shows the plots of $T_m$ vs. 
$H^{2/3}$ for samples 1, 2, 3, 5, and 6. 
$T_g$ is the intercept of the best-fitted straight line with the $T_m$-axis, i.e., 
$T_m \rightarrow T_g$ when $H \rightarrow 0$.
This $T_g$ is also pronounced in $\chi_{ac}$
measurements, i.e., we got a peak in ac-magnetization 
measurements at roughly the same
temperature as shown in Fig. \ref{fig:s2ac} and 
discussed in subsection~\ref{ss:acm}.

\begin{figure}
\includegraphics[bb=0.4cm 0.5cm 10.5cm 7.8cm,width=8.5cm]{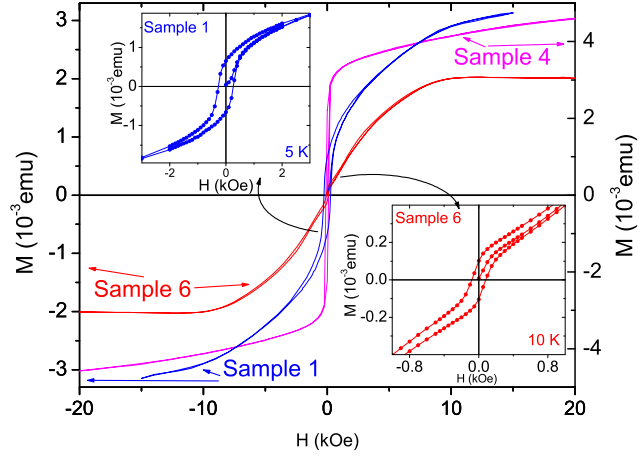}
\caption{\label{fig:mh} (Color online) Magnetization {\it vs.} external magnetic field
for samples 1 at 5 K, 4 at 10 K, and 6 at 10 K. Insets are the same M(H) 
plots for samples 1 and 6 but on different expanded scales.}
\end{figure}

\begin{figure}
\includegraphics[bb=0.4cm 0.5cm 10.5cm 7.8cm,width=8.5cm]{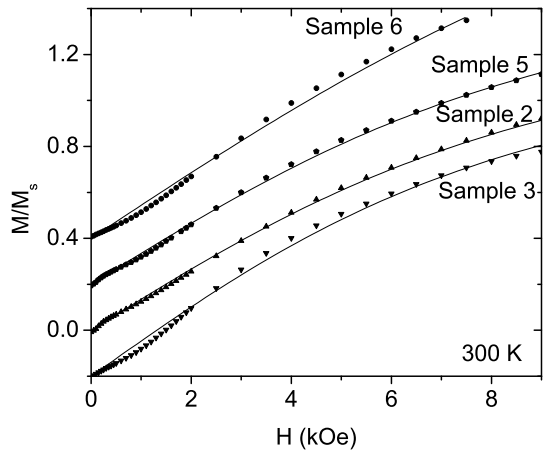}
\caption{\label{fig:lang} The Langevin function fits for samples 2, 3,
5, and 6 below $H_{sat}$. Data and fit for samples 3, 5, and 6 are 
shifted along y-axis by adding -0.2, 0.2, and 0.4, respectively for clarity.}
\end{figure}

\subsection{M-H loop}

m-h measurements of these samples are shown in Fig. \ref{fig:mh}. 
All the samples except 4 show typical hysteresis of AF coupled multilayers.
From these measurements we
are able to extract the following important information:

\begin{enumerate}
\item {\it B series} samples have better antiferromagnetic coupling (except sample 4) between 
Fe layers compared to {\it A series} samples. We have made a comparison between samples 
according to their remanent to saturation magnetization ratio $(M_r/M_s)$. 
We define a quantity called the antiferromagnetic fraction(AFF) as

\begin{equation}
AFF (\%) = \left(1 - \frac{M_r}{M_s} \right) \times 100 \%. 
\end{equation}

Samples 1, 2, and 3 have AFF $\sim$ 80 \% at 5 K. Sample 4, which has $<$ 1 \% GMR at 5 K has 
FM alignment of Fe layers even in zero applied magnetic field, has the least AFF $\sim$ 50 \%.
Samples 5, 6, 7, and 8 have AFF $\sim$ 95 \% at 10 K. This gives us an important information about
the coupling of Fe layers. Fe layers in samples 5, 6, 7, and 8 are coupled antiferromagnetically
better than samples 1, 2, and 3 in zero field. Samples 1, 2, and 3 have more interface roughness 
compared to sample 4, 5, 6, 7, and 8 (concluded from the fact that samples 
5, 6, 7, and 8 have less residual resistivity ($\sim$ 30 $\mu \Omega$cm) compared to 
samples 1, 2, and 3 ($\sim$ 45 $\mu \Omega$cm). The
saturation field for {\it A series} samples is about 1.3 tesla but the saturation 
fields for for {\it B series} samples decreases gradually from 1.2 to 0.5 tesla.
This implies that the strength of coupling in {\it A series} samples
is stronger than {\it B series} samples.
For {\it A series} samples the coercive field is about 200 Oe at 10 K. 
The coercive field of samples 5 - 8 is about 70 Oe. This lower value of the
coercive field indicates that samples 5 - 8 have ``cleaner'' interfaces,
i. e., in the absence of a field or in small fields the Fe layers are aligned ``perfectly'' AF. Samples 1, 
2, and 3 have ``rougher'' interfaces and there are more Fe cluster embedded in Cr layer
and vice versa. Probably this is one of the reasons why the history-dependent behavior
was more prominently seen in samples 1 - 3. The MBE grown samples,
however, have never shown any hysteresis effect due to their much smoother interfaces.
 
 \item The magnetization data with external applied magnetic field at 
 room temperature fit well to the Langevin function of the form
 \begin{equation}
 \label{eq:lang}
 M(H)=N \bar{\mu} \left[ \coth \left( \frac{\bar{\mu}H}{kT} \right) - \frac{kT}{\bar{\mu}H} \right],
 \end{equation}
 where $\bar{\mu}$ is the average magnetic moment of the clusters, N is the number 
of clusters in the sample and $N\bar{\mu}$ is the amplitude of the superparamagnetic 
contribution. Figure~\ref{fig:lang} shows the Langevin function fit for samples 2, 3, 5, 6, and 7
below saturation field. The average magnetic moment per superparamagnetic cluster
$\bar{\mu}$ in Bohr magnetons have been found $\sim 1200 \pm 20, 1280 \pm 70,
1200 \pm 30, 825 \pm 100,$ and $1150 \pm 230 \ \mu_B$ 
for samples 2, 3, 5, 6, and 7, respectively. This gives
an indication that the average interdiffused particle volume is about the same in both
the series and history dependent behaviour is affected prominently by 
interface heterostructure. We did not get Langevin function fit for sample 8
which has the least saturation field of about half a tesla. Here, the
exchange coupling between successive Fe layers are weak but the 
interdiffused clusters are strongly coupled to the nearby Fe layers.

\end{enumerate}

\subsection{ \label{ss:acm} ac-magnetization}
\begin{figure}
\includegraphics[bb=0.4cm 0.5cm 9.5cm 7.2cm,width=8.5cm]{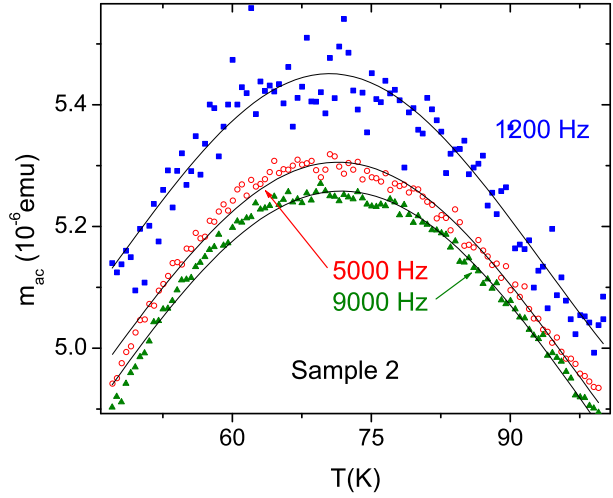}
\caption{\label{fig:s2ac} (Color online) ac-magnetization {\it vs.} temperature for
sample 2 at different frequencies. The points represent the data 
whereas the solid lines are the fits to find $T_g$.}
\end{figure}

\begin{figure}
\includegraphics[bb=0.4cm 0.5cm 9.5cm 7.2cm,width=8.5cm]{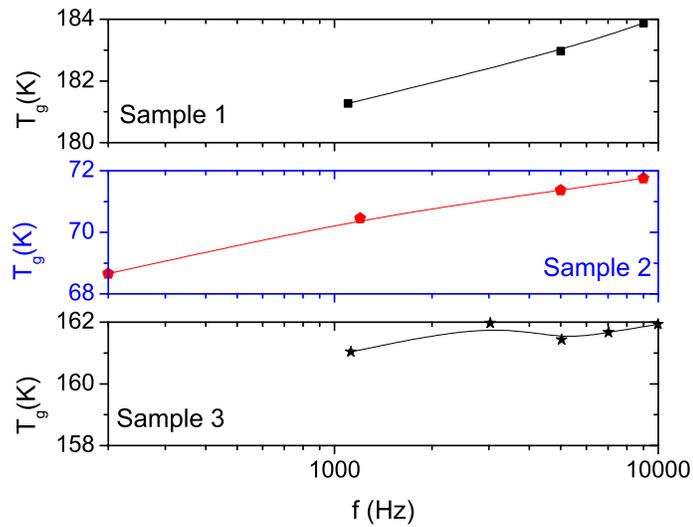}
\caption{\label{fig:tg} (Color online) Variation of peak temperature ($T_g$) with frequency
of the applied magnetic field. The solid lines are just guides to the eye.}
\end{figure}

We have measured the ac-magnetization vs. temperature in these multilayers at  
different frequencies of an ac field of 10 Oe. We have found that the ac-magnetization 
has peaks as shown in Fig.~\ref{fig:s2ac} for sample 2. As the signals are quite low (micro-gram of Fe in these samples) we got rather noisy data.
Here the peaks in the ac-magnetization show a temperature shift, although by a 
very small amount (as in spin glasses), when we change the frequency of the driving signal. 
For comparison, a quantitative measure of the frequency shift in terms of $(\Delta T_g/T_g)\times 100 \%$
per decade of $\omega$ in canonical spin-glass is $Cu$Mn (0.5 \%), $Au$Mn (0.45 \%), and $Ag$Mn
(0.6 \%)~\cite{Mydosh:1993}. For the insulating spin glasses the frequency dependence is larger.
It is much more in superparamagnets like $a-CoO.Al_2O_3SiO_2$ (6 \%) 
and $a-(Ho_2O_3)(B_2O_3)$ (28 \%).
Spin glasses show cusp-like behavior at $T_g$. The cusp gets smeared even at 
fields only as high as 50 Oe. In our measurements we found only rounded peaks.
The peak shifts to higher temperatures as we increase the frequency.
To analyse this data we did peak fitting to each data set. From this peak fit we are able to
observe clearly that the peak shifts toward higher temperature as shown in Fig.~\ref{fig:tg}
and the magnitude of the magnetization decreases with increasing frequency. 
To interpret this result we assume the presence of clusters of different volumes 
(giving rise to a distribution of moments) at the interfaces and/or interdiffused in the bulk. 
At lower frequency moments of all size will be able to follow the magnetic field while 
at larger frequencies only small moments will be following the external field and the 
large clusters (large moments) will not be able to respond. This explains the larger 
magnitude of magnetization for lower frequency. Any peak is a competition of two 
processes. At the lowest temperature the moments are frozen but
as we start raising the temperature the thermal energy gives them 
some freedom to move and we get higher magnetization with increasing temperature. 
Above a certain temperature there is no freezing and we get a maximum. Further 
increase in the thermal energy disrupts all the alignments and so the moment
starts falling. Clusters of smaller size are unlocked at lower temperatures 
compared to those having large moments. The Arrhenius law to
explain the frequency shift in superparamagnets 
is given by $\omega = \omega_0 \ exp[-E_a/k_BT_g]$, where $\omega$
is the driving frequency of $\chi_{ac}$ measurements, $E_a$ is the energy
barrier height (=$KV$, the anisotropy constant times the cluster volume), and 
$T_g$ is the peak temperature. If the volume of clusters is fixed, then
$T_g$ increases with increasing frequency. However, if there is a
cluster volume distribution, then 
at higher frequency only smaller clusters respond and so the peak temperature 
shifts towards lower temperatures with increasing frequency. 
In multilayers, $T_g$ is shifting towards higher temperatures with
increasing frequency. This suggests that the change in $T_g$ 
due to cluster volume change is small in the present investigation.

To conclude, the low-field magnetization behavior has been studied in ion-beam sputtered Fe-Cr GMR
multilayers. Magnetic heterostructure due to the interdiffused particles/cluster and the interfacial
imperfactions play an important role in the low-field magnetization behavior. In low fields and at low 
temperatures the presence of spin-glass-like phase is established. At higher temperatures these 
multilayers behave more like a superparamagnet.

\begin{acknowledgments}
One of us (R.S.P.) acknowledges CSIR, Government of India for financial support. We thank Dr.
Ashna Bajpai for help and encouragement during some measurements. We sincerely thank
Dr. R. N. Viswanath of Forschungszentrum, Karlsruhe for doing the ac-magnetization measurements.
\end{acknowledgments}



\begin{table*}
\caption{\label{tab:dct} Three characterstic temperatures $T_m$, $T_{irr}$, and
$T_{inf}$ as functions of the applied magnetic field.}
\begin{tabular}{|c|c|c|c|c|c|c|}
\hline
\emph{Characterstic} & \multicolumn{6}{c|}{\emph{Applied magnetic field}} \\ \cline{2-7}
\emph{Temperatures} & 10 Oe & 50 Oe & 100 Oe & 150 Oe & 200 Oe & Sample \\ \hline
   & 135 & 150 & 123 & 105 & 100 & 1   \\ \cline{2-7}
$T_m$ (K) & 41 & 41 & 35 & 29 & 24 & 2  \\ \cline{2-7}
($\pm$ 2 K) & 95 & 97 & 83 & 74 & 60 & 3  \\ \cline{2-7}
& NP\footnotemark[1] &  & NP &  & NP  & 4  \\ \cline{2-7}
&  & 75 & 62 & 58 & 58 & 5  \\ \cline{2-7}
& 30 & 20 & 14 &  & 8 & 6  \\ \cline{2-7}
&  & 8 &  &  & NP & 7  \\ \cline{2-7}
   &  & NP & NP &  & NP & 8  \\ \hline
 &  & 140 & 140 & 135 & 140 & 1  \\ \cline{2-7}
$T_{inf}$ (K) & 43 & 45 & 45 & 43 & 47 & 2 \\ \cline{2-7}
($\pm$ 5 K) & 91 & 97 & 97 & 98 & 98 &  3 \\ \cline{2-7}
   & NP &  & NP &  & NP & 4  \\ \cline{2-7}
&  & 75 & 75 & 75 & 85 & 5  \\ \cline{2-7}
& 20 & 22 & 20 &  & 22 & 6  \\ \cline{2-7}
&  & NP &  &  & NP & 7  \\ \cline{2-7}
   &  & NP & NP &  & NP & 8  \\ \hline
 &  & 290 & 245 & 225 & 220 & 1  \\ \cline{2-7}
$T_{irr} (K) $ & 191 & 175 & 149 & 139 & 210  & 2  \\ \cline{2-7}
($\pm$ 5 K) & 265 & 230 & 205 & 200 & 195 & 3  \\ \cline{2-7}
   & NP &  & NP &  & NP & 4  \\ \cline{2-7}
&  & 280 & 200 & 200 & 195 & 5  \\ \cline{2-7}
& 245 & 210 & 200 &  & 145 & 6  \\ \cline{2-7}
&  & 100 &  &  & 200 & 7  \\ \cline{2-7}
   &  & NP & NP &  & NP & 8  \\ \hline
\end{tabular}
\footnotetext[1]{NP: Feature Not Present}
\end{table*}

\begin{table}
\caption{Values of $\chi^2$, correlation coefficient 
R$^2$, the parameters $M(0)$, and $A$ of Eq. (~\ref{eq:pwr}).}
\begin{tabular}{ccccccc} \hline
H(Oe)  &  $\chi^2(10^{-6})\footnotemark[2]$  &  $R^2$  &  $M(0)(emu/cc)$      &  A$(10^{-4}K^{-3/2})$
     &  Range of T(K)   &  \# of data points \\ \hline \hline
{\it Sample 1}  &    &    &    &    &    &   \\  
50  &  3.984  &  0.9998  &  138.0$\pm$0.1  &  2.422$\pm$0.003  &  5 - 140  &  68 \\  
100  &  2.242  &  0.9999  &  227.9$\pm$0.1  &  2.467$\pm$0.003  &  5 - 140  &  68 \\  
150  &  10.807  &  0.9995  &  353.9$\pm$0.2  & 2.570$\pm$0.008  &  10 - 140  &  41 \\  
200  &  11.037  &  0.9991  &  390.7$\pm$0.2  &  2.351$\pm$0.007  &  5 - 140  &  60 \\
13000\footnotemark[3]  &  0.005  &  0.9901  &  1284.3$\pm$0.1  &  0.101$\pm$0.001  &  5 - 100  &  95 \\ \hline 

{\it Sample 3}  &    &    &    &    &        &  \\  
10  &  8.345  &  0.9990  &  45.63$\pm$0.03  &  2.63$\pm$0.01  &  5 - 100  &  47 \\  
50  &  19.325  &  0.9985  &  130.7$\pm$0.1  &  3.28$\pm$0.02  &  5 - 100  &  47 \\  
100  &  32.110  &  0.9978  &  199.4$\pm$0.2  &  3.40$\pm$0.02  &  5 - 100  &  47 \\  
150  &  6.232  &  0.9995  &  269.3$\pm$0.2  &  3.27$\pm$0.01  &  5 - 100  &  46 \\  
200  &  7.480  &  0.9994  &  303.2$\pm$0.2  &  3.19$\pm$0.01  &  5 - 100  &  46 \\ 
13000\footnotemark[3]  &  0.006  &  0.9963  &  1463.0$\pm$0.1  &  0.100$\pm$0.001  &  5 - 100  &  95 \\ \hline 

{\it Sample 4}  &    &    &    &    &       &  \\  
10 (FC) &  1.066  &  0.9992  &  1105.4$\pm$0.2  &  0.218$\pm$0.001  &  5 - 300  &  114 \\
10 (ZFC) &  1.359  &  0.9990  &  1106.4$\pm$0.2  &  0.221$\pm$0.001  &  5 - 300  &  114 \\
100 (FC) &  1.065  &  0.9977  &  1459.0$\pm$0.3  &  0.130$\pm$0.001  &  5 - 300  &  60 \\  
100 (ZFC) &  1.448  &  0.9972  &  1463.1$\pm$0.4  &  0.137$\pm$0.001  &  5 - 300  &  60 \\ \hline

{\it Sample 5}  &    &    &    &    &     &  \\  
50  & 0.608   & 0.9996 & 138.6$\pm$0.1 & 3.42$\pm$0.01 & 5 - 80 & 23  \\
100 & 65.962  & 0.9976 & 190.4$\pm$0.4 & 4.48$\pm$0.03 & 5 - 80 & 26  \\  
150 & 37.940  & 0.9983 & 243.5$\pm$0.3 & 2.27$\pm$0.02 & 5 - 80 & 23  \\
200 & 17.731  & 0.9979 & 356.4$\pm$0.2 & 3.82$\pm$0.02 & 5 - 80 & 76 \\ \hline \hline

\end{tabular}
\label{tab:pwr}
\footnotetext[2]{ $\chi^2$ is defined as 
\begin{equation}
\chi^2 = \frac{1}{N} \sum_{i=1}^{N} 
\frac{(M_{i-measured}-M_{i-fitted})^2}{M_{i-mean}^2} \nonumber
\end{equation}
}
\footnotetext[3]{Ref.~\onlinecite{Patel:2005}}
\end{table}

\end{document}